\documentclass[preprint,amsmath,amssymb,aps,showkeys,showpacs]{revtex4}
\usepackage[english]{babel}
\usepackage{graphicx}
\usepackage{graphics}
\usepackage{amsmath}
\usepackage{dcolumn}
\usepackage{amssymb}
\usepackage{bm}
\usepackage[latin1]{inputenc}

\begin{document}
\title{On the Klein-Gordon oscillator subject to a Coulomb-type potential}
\author{K. Bakke}
\email{kbakke@fisica.ufpb.br} 
\affiliation{Departamento de F\'isica, Universidade Federal da Para\'iba, Caixa Postal 5008, 58051-970, Jo\~ao Pessoa, PB, Brazil.} 
\author{C. Furtado}
\email{furtado@fisica.ufpb.br} 
\affiliation{Departamento de F\'isica, Universidade Federal da Para\'iba, Caixa Postal 5008, 58051-970, Jo\~ao Pessoa, PB, Brazil.}

\begin{abstract}
By introducing the scalar potential as modification in the mass term of the Klein-Gordon equation, the influence of a Coulomb-type potential on the Klein-Gordon oscillator is investigated. Relativistic bound states solutions are achieved to both attractive and repulsive Coulomb-type potentials and the arising of a quantum effect characterized by the dependence of angular frequency of the Klein-Gordon oscillator on the quantum numbers of the system is shown. 
\end{abstract}


\maketitle

\section{Introduction}

In recent decades, the relativistic generalization of the harmonic oscillator has attracted a great deal of attention. The most known relativistic model of the harmonic oscillator was introduced by Moshinsky and Szczepaniak \cite{osc1}, which is known as the Dirac oscillator, and has been investigated by several authors \cite{osc2,osc2a,osc3,b11,bf40,josevi,jay2,osc6,extra2,extra3}. In particular, the Dirac oscillator has attracted a great interest in studies of Jaynes-Cummings model \cite{jay2,osc3}, the Ramsey-interferometry effect \cite{osc6} and quantum phase transitions \cite{extra2,extra3}. On the other hand, a relativistic model of the harmonic oscillator has also been proposed for a scalar particle by Bruce and Minning \cite{kgo} based on the Dirac oscillator \cite{osc1}. Bruce and Minning \cite{osc1} showed that an analogous coupling to the linear Dirac oscillator coupling can be introduced into the Klein-Gordon equation in such a way that one can recover the Schr\"odinger equation for a harmonic oscillator in the nonrelativistic limit. This coupling proposed by Bruce and Minning is known as the Klein-Gordon oscillator \cite{kgo,kgo2,kgo8,kgo7}. As example, by considering the isotropic Klein-Gordon oscillator in $\left(2+1\right)$ dimensions, the Klein-Gordon equation becomes:
\begin{eqnarray}
\left[\mathcal{E}^{2}-m^{2}\right]\phi=\left[\hat{p}+im\omega\rho\,\hat{\rho}\right]\cdot\left[\hat{p}-im\omega\rho\,\hat{\rho}\right]\phi,
\label{kgo}
\end{eqnarray}
where $m$ is the rest mass of the scalar particle, $\omega$ is the angular frequency of the Klein-Gordon oscillator, $\rho=\sqrt{x^{2}+y^{2}}$ and $\hat{\rho}$ is a unit vector in the radial direction. In recent years, the Klein-Gordon oscillator has been investigated in noncommutative space \cite{kgo3,kgo4}, in noncommutative phase space \cite{kgo5} and in $\mathcal{PT}$-symmetric Hamiltonian \cite{kgo6}.

Recently, several authors have shown the interest in investigating relativistic effects \cite{bah,bah2,bah3,bah4} on systems where the motion of a particle is governed by harmonic oscillations, such as the vibrational spectrum of diatomic molecules \cite{ikh}, the binding of heavy quarks \cite{qui,chai} and the oscillations of atoms in crystal lattices, by mapping them as a position-dependent mass system \cite{pdm,pdm2,pdm3,pdm4}. The importance of these potentials arises from the presence of a strong potential field. In particular, Bahar and Yasuk \cite{bah} dealt with the quark-antiquark interaction as a problem of a relativistic spin-$0$ particle possessing a position-dependent mass, where the mass term acquires a contribution given by a interaction potential that consists of a linear and a harmonic confining potential plus a Coulomb potential term. It is worth mentioning other works that have explored the relativistic quantum dynamics of a scalar particle subject to different confining potentials which can be in the interest of several areas of physics \cite{alvaro,qian,castro,alhardi,adame,xu}.

The aim of this work is to study the influence of a Coulomb-type potential on the Klein-Gordon oscillator. In recent years, the confinement of a relativistic scalar particle to a Coulomb potential has been discussed by several authors \cite{kg,kg2,kg3,kg4,greiner}. As discussed in Ref. \cite{greiner}, the procedure in introducing a scalar potential into the Klein-Gordon equation follows the same procedure in introducing the electromagnetic 4-vector potential. This occurs by modifying the momentum operator $p_{\mu}=i\partial_{\mu}$ in the form: $p_{\mu}\rightarrow p_{\mu}-q\,A_{\mu}\left(x\right)$. Another procedure was proposed in Ref. \cite{scalar} by making a modification in the mass term in the form: $m\rightarrow m+S\left(\vec{r},\,t\right)$, where $S\left(\vec{r},\,t\right)$ is the scalar potential. This modification in the mass term has been explored in recent decades, for instance, by analysing the behaviour of a Dirac particle in the presence of static scalar potential and a Coulomb potential \cite{scalar2} and a relativistic scalar particle in the cosmic string spacetime \cite{eug}. In this work, we investigate the influence of a Coulomb-type potential on the Klein-Gordon oscillator by introducing the scalar potential as modification in the mass term in the Klein-Gordon equation. We obtain bound state solutions to the Klein-Gordon equation for both attractive and repulsive Coulomb-type potentials and show a quantum effect characterized by the dependence of angular frequency of the Klein-Gordon oscillator on the quantum numbers of the system, which means that not all values of the angular frequency are allowed.

The structure of this paper is as follows: in section II, we study the Klein-Gordon oscillator subject to a Coulomb-type potential in the Minkowski spacetime in $\left(2+1\right)$ dimensions; in section III, we present our conclusions.

\section{Klein-Gordon oscillator under the influence of a Coulomb-type potential}

In this section, we study the behaviour of the Klein-Gordon oscillator subject to a Coulomb-type potential in (2+1) dimensions. We consider the cylindrical symmetry, then, we write the line element of the Minkowski spacetime in the form (with $c=\hbar=1$):
\begin{eqnarray}
ds^{2}=-dt^{2}+d\rho^{2}+\rho^{2}\,d\varphi^{2}.
\label{1}
\end{eqnarray}

Thereby, the Klein-Gordon equation describing the interaction of the Klein-Gordon oscillator and static scalar potential is given by  (with $c=\hbar=1$)
\begin{eqnarray}
\left[m+S\left(\rho\right)\right]^{2}\phi=-\frac{\partial^{2}\phi}{\partial t^{2}}-\left[\hat{p}+im\omega\rho\,\hat{\rho}\right]\cdot\left[\hat{p}-im\omega\rho\,\hat{\rho}\right]\phi,
\label{2}
\end{eqnarray}
where $S\left(\rho\right)$ is a scalar potential, $m$ is the rest mass of the scalar particle, $\omega$ is the angular frequency of the Klein-Gordon oscillator, $\hat{p}=-i\vec{\nabla}$, $\rho=\sqrt{x^{2}+y^{2}}$ and $\hat{\rho}$ is a unit vector in the radial direction. In this way, by considering a Coulomb-type potential 
\begin{eqnarray}
S\left(\rho\right)=\frac{f}{\rho}=\pm\frac{\left|f\right|}{\rho},
\label{3}
\end{eqnarray}
where $f$ is a constant, the Klein-Gordon equation (\ref{2}) becomes
\begin{eqnarray}
-\frac{\partial^{2}\phi}{\partial t^{2}}+\frac{\partial^{2}\phi}{\partial\rho^{2}}+\frac{1}{\rho}\frac{\partial\phi}{\partial\rho}+\frac{1}{\rho^{2}}\frac{\partial^{2}\phi}{\partial\varphi^{2}}+m\omega\,\phi-m^{2}\omega^{2}\rho^{2}\,\phi-m^{2}\phi-\frac{2mf}{\rho}\,\phi-\frac{f^{2}}{\rho^{2}}\,\phi=0.
\label{4}
\end{eqnarray}

In what follows, we shall be considering particular stationary solutions to Eq. (\ref{4}) that are eigenfunctions of the operator $\hat{L}_{z}=-i\partial_{\varphi}$. In this way, we have
\begin{eqnarray}
\phi=e^{-i\mathcal{E}t}\,e^{il\varphi}\,R\left(\rho\right),
\label{5}
\end{eqnarray}
 where $l=0,\pm1,\pm2,\ldots$. Then, substituting (\ref{5}) into Eq. (\ref{4}), we obtain
\begin{eqnarray}
\frac{d^{2}R}{d\rho^{2}}+\frac{1}{\rho}\frac{dR}{d\rho}-\frac{\gamma^{2}}{\rho^{2}}\,R-\frac{2mf}{\rho}\,R-m^{2}\omega^{2}\rho^{2}\,R+\beta^{2}R=0,
\label{6}
\end{eqnarray}
where we have defined the following parameters in Eq. (\ref{6}):
\begin{eqnarray}
\beta^{2}&=&\mathcal{E}^{2}-m^{2}+m\omega;\nonumber\\
[-2mm]\label{7}\\[-2mm]
\gamma^{2}&=&l^{2}+f^{2}.\nonumber
\end{eqnarray}

From now on, let us consider $\xi=\sqrt{m\omega}\,\rho$, thus, we rewrite the radial equation (\ref{6}) in the form:
\begin{eqnarray}
\frac{d^{2}R}{d\xi^{2}}+\frac{1}{\xi}\frac{dR}{d\xi}-\frac{\gamma^{2}}{\xi^{2}}\,R-\frac{\delta}{\xi}\,R-\xi^{2}\,R+\frac{\beta^{2}}{m\omega}\,R=0,
\label{8}
\end{eqnarray}
where we have defined a new parameter
\begin{eqnarray}
\delta=\frac{2mf}{\sqrt{m\omega}}.
\label{9}
\end{eqnarray}

Let us discuss the asymptotic behaviour of the possible solutions to Eq. (\ref{8}), which are determined for $\xi\rightarrow0$ and $\xi\rightarrow\infty$. From Refs. \cite{eug,mhv,vercin}, the behaviour of the possible solutions to Eq. (\ref{8}) at $\xi\rightarrow0$ and $\xi\rightarrow\infty$ allows us to write the function $R\left(\xi\right)$ in terms of an unknown function $H\left(\xi\right)$ as it follows:
\begin{eqnarray}
R\left(\xi\right)=\exp\left(-\frac{\xi^{2}}{2}\right)\,\xi^{\sigma\left|\gamma\right|}\,H\left(\xi\right),
\label{10}
\end{eqnarray}
where $\sigma =\pm 1$. Note that one should have in mind the possibility of existing a singular solution to Eq. (\ref{8}). As pointed out in Refs. \cite{mhv,sin16}, $\sigma=1$ yields a non-singular solution whereas $\sigma=-1$ yields a singular solution to Eq. (\ref{8}). In this contribution, we consider only non-singular solutions to Eq.(\ref{8}), that is, $\sigma=1$. Then, substituting the radial wave function given in Eq. (\ref{10}) into Eq. (\ref{8}), we obtain
\begin{eqnarray}
\frac{d^{2}H}{d\xi^{2}}+\left[\left(2\left|\gamma\right|+1\right)\frac{1}{\xi}-2\xi\right]\frac{dH}{d\xi}+\left[\frac{\beta^{2}}{m\omega}-2-2\left|\gamma\right|-\frac{\delta}{\xi}\right]H=0.
\label{11}
\end{eqnarray}

The second order differential equation (\ref{11}) corresponds to the Heun biconfluent equation \cite{heun,eug,bm,b50} and the function $H\left(\xi\right)$ is the Heun biconfluent function
\begin{eqnarray}
H\left(\xi\right)=H\left(2\left|\gamma\right|,\,0,\,\frac{\beta^{2}}{m\omega},\,2\delta,\xi\right).
\label{12}
\end{eqnarray}

Proceeding with our discussion about bound states solutions, let us use the Frobenius method \cite{arf,f1}. Thereby, the solution to Eq. (\ref{11}) can be written as a power series expansion around the origin:
\begin{eqnarray}
H\left(\xi\right)=\sum_{j=0}^{\infty}\,a_{j}\,\xi^{j}.
\label{13}
\end{eqnarray} 

Substituting the series (\ref{13}) into (\ref{1}), we obtain the following recurrence relation:
\begin{eqnarray}
a_{j+2}=\frac{\delta}{\left(j+2\right)\,\left(j+1+\alpha\right)}\,a_{j+1}-\frac{\left(\theta-2j\right)}{\left(j+2\right)\,\left(j+1+\alpha\right)}\,a_{j},
\label{14}
\end{eqnarray}
where $\alpha=2\left|\gamma\right|+1$ and $\theta=\frac{\beta^{2}}{m\omega}-2-2\left|\gamma\right|$. By starting with $a_{0}=1$ and using the relation (\ref{14}), we can calculate other coefficients of the power series expansion (\ref{13}). For instance,
\begin{eqnarray}
a_{1}&=&\frac{\delta}{\alpha}\nonumber\\
&=&\frac{2mf}{\sqrt{m\omega}}\frac{1}{\left(2\left|\gamma\right|+1\right)};\nonumber\\
[-2mm]\label{15}\\[-2mm]
a_{2}&=&\frac{\delta^{2}}{2\alpha\left(1+\alpha\right)}-\frac{\theta}{2\left(1+\alpha\right)}\nonumber\\
&=&\frac{2mf^{2}}{\omega}\frac{1}{\left(2\left|\gamma\right|+1\right)\left(2\left|\gamma\right|+2\right)}-\frac{\theta}{2\left(2\left|\gamma\right|+2\right)}.\nonumber
\end{eqnarray}

It is well-known that the quantum theory requires that the wave function (\ref{5}) must be normalizable. Therefore, we assume that the function $R\left(\xi\right)$ vanishes at $\xi\rightarrow0$ and $\xi\rightarrow\infty$. This means that we have a finite wave function everywhere, that is, there is no divergence of the wave function at $\xi\rightarrow0$ and $\xi\rightarrow\infty$, then, bound state solutions can be obtained. However, we have written the function $H\left(\xi\right)$ as a power series expansion around the origin in Eq. (\ref{13}). Thereby, bound state solutions can be achieved by imposing that the power series expansion (\ref{13}) or the Heun biconfluent series becomes a polynomial of degree $n$. In this way, we guarantee that $R\left(\xi\right)$ behaves as $\xi^{\left|\gamma\right|}$ at the origin and vanishes at $\xi\rightarrow\infty$ \cite{vercin,mhv}. Through the recurrence relation (\ref{14}), we can see that the power series expansion (\ref{13}) becomes a polynomial of degree $n$ by imposing two conditions \cite{f1,bb2,bb4,bm,eug,b50,vercin,mhv}:
\begin{eqnarray}
\theta=2n\,\,\,\,\,\,\mathrm{and}\,\,\,\,\,\,a_{n+1}=0,
\label{16}
\end{eqnarray}
where $n=1,2,3,\ldots$. From the condition $\theta=2n$, we can obtain:
\begin{eqnarray}
\mathcal{E}_{n,\,l}^{2}=m^{2}+2m\,\omega_{n,\,l}\left[n+\left|\gamma\right|+\frac{1}{2}\right].
\label{16a}
\end{eqnarray}

Hence, Eq. (\ref{16a}) is a general expression for the relativistic energy levels of the Klein-Gordon oscillator subject to a Coulomb-type potential, where we have coupled this scalar potential as a modification of the mass term in the Klein-Gordon equation. In contrast to the results of Ref. \cite{kgo}, we have that the influence of the Coulomb-like potential makes that the ground state to be defined by the quantum number $n=1$ instead of the quantum number $n=0$. Note that we have written the angular frequency $\omega$ in terms of the quantum numbers $\left\{n,\,l\right\}$ in Eq. (\ref{16a}). From the mathematical point of view, this dependence of the angular frequency of this relativistic oscillator on the quantum numbers $\left\{n,\,l\right\}$ results from the fact that the exact solutions to Eq. (\ref{11}) are achieved for some values of the Klein-Gordon oscillator frequency. From the quantum mechanics point of view, this is an effect which arises from the influence of the Coulomb-type potential on the Klein-Gordon oscillator.

Henceforth, let us discuss this behaviour of the Klein-Gordon oscillator frequency. First of all, we must note that, at first glance, the result obtained in Eq. (\ref{16a}) shows that the relativistic energy levels do not depend on the parameter $\delta$ which gives rise to the Coulomb-type potential (defined in Eq. (\ref{9})). However, we need to analyse the second condition established in Eq. (\ref{16}), that is, the condition $a_{n+1}=0$. By analysing this second condition, we obtain an expression involving specific values of the angular frequency of the Klein-Gordon oscillator and the parameter $\delta$. It is worth mentioning other analyses which have been made in recent years, as example, a relation involving the harmonic oscillator frequency, the Lorentz symmetry-breaking parameters and the total angular momentum quantum number obtained in Ref. \cite{bb2}, a relation involving a coupling constant of a Coulomb-like potential, the cyclotron frequency and the total angular momentum quantum number in semiconductors threaded by a dislocation density obtained in Ref. \cite{bm}, and a relation involving the mass of a relativistic particle, a scalar potential coupling constant and the total angular momentum quantum number achieved in Ref. \cite{eug}.

Thereby, let us assume that the angular frequency of the Klein-Gordon oscillator $\omega$ can be adjusted in such a way that the condition $a_{n+1}=0$ is satisfied in order that a relation between the angular frequency of the Klein-Gordon oscillator and the Coulomb-type potential parameter can be obtained. The meaning of achieving this relation is that not all values of the angular frequency $\omega$ are allowed, but some specific values of $\omega$ which depend on the quantum numbers $\left\{n,\,l\right\}$. For this reason, we label 
\begin{eqnarray}
\omega=\omega_{n,\,l}. 
\label{16b}
\end{eqnarray}
Therefore, the conditions established in Eq. (\ref{16}) are satisfied and a polynomial solution to the function $H\left(\xi\right)$ given in Eq. (\ref{13}) is achieved \cite{vercin,mhv,eug,b50}. As an example, let us consider $n=1$, which corresponds to the ground state, and analyse the condition $a_{n+1}=0$. For $n=1$, we have $a_{2}=0$. The condition $a_{2}=0$, thus, yields 
\begin{eqnarray}
\omega_{1,\,\,l}=\frac{2mf^{2}}{\left(2\left|\gamma\right|+1\right)}.
\label{17}
\end{eqnarray}

In this way, substituting Eq. (\ref{17}) into Eq. (\ref{16}), the expression of the energy level ( for $n=1$) becomes:
\begin{eqnarray}
\mathcal{E}_{1,\,l}&=&\pm\sqrt{m^{2}+2m\omega_{1,\,l}\left[n+\left|\gamma\right|+\frac{1}{2}\right]}\nonumber\\
[-2mm]\label{18}\\[-2mm]
&=&\pm m\left[1+4f^{2}\frac{\left(\left|\gamma\right|+\frac{3}{2}\right)}{\left(2\left|\gamma\right|+1\right)}\right]^{1/2}.\nonumber
\end{eqnarray}

In what follows, let us consider the simplest case of the function $H\left(\xi\right)$ which corresponds to a polynomial of first degree. In this way, for $n=1$, we can write
\begin{eqnarray}
H_{1,\,l}\left(\xi\right)=1+\frac{\delta}{\alpha}\,\xi.
\label{19}
\end{eqnarray}
Thereby, from Eq. (\ref{19}), thus, Eq. (\ref{10}) can be written in the form: $R_{1,\,l}\left(\xi\right)=e^{-\xi^{2}/2}\,\xi^{\left|\gamma\right|}\,\left(1+\frac{\delta}{\alpha}\,\xi\right)$.

Hence, the effects of the Coulomb-type potential introduced by a coupling with the mass term on the spectrum of energy of the Klein-Gordon oscillator correspond to a change of the energy levels, where the ground state is defined by the quantum number $n=1$. Moreover, a quantum effect characterized by the dependence of angular frequency of the Klein-Gordon oscillator on the quantum numbers $\left\{n,l\right\}$ of the system arises, whose meaning is that not all values of the angular frequency are allowed. In this way, the conditions established in Eq. (\ref{16}) are satisfied and a polynomial solution to the function $H\left(\xi\right)$ given in Eq. (\ref{13}) is achieved in agreement with Refs. \cite{vercin,mhv,eug,b50}.

\section{Conclusions}

We have studied the influence of a Coulomb-type potential on the Klein-Gordon oscillator and shown that the ground state of the Klein-Gordon oscillator is defined by the quantum number $n=1$ instead of the quantum number $n=0$. Other quantum effect obtained is the dependence of angular frequency of the Klein-Gordon oscillator on the quantum numbers $\left\{n,l\right\}$ of the system, whose meaning is that not all values of the angular frequency are allowed. As an example, we have calculated the angular frequency of the ground state $n=1$ and obtained the expression of the energy level of the ground state.

It is worth mentioning that we have introduced the scalar potential as a modification of the mass term of the Klein-Gordon equation. However, as discussed in Ref. \cite{greiner}, one can use the same procedure of introducing the electromagnetic 4-vector potential by modifying the momentum operator as $p_{\mu}\rightarrow p_{\mu}-q\,A_{\mu}\left(x\right)$. By dealing with a Coulomb-type potential via a minimal coupling, new interesting results associated with the Klein-Gordon oscillator can be obtained. Moreover, the interaction between a relativistic scalar particle and the Klein-Gordon oscillator can be of interest in studies of the quark-antiquark interaction \cite{bah}, position-dependent mass systems \cite{pdm,pdm2,pdm3,pdm4}, the Casimir effect \cite{casimir,mb1} and the Kaluza-Klein theory \cite{furtado}.

\acknowledgments{The authors would like to thank CNPq (Conselho Nacional de Desenvolvimento Cient\'ifico e Tecnol\'ogico - Brazil) for financial support. }

\end{document}